\documentclass[pra,aps]{revtex4}
\usepackage{graphicx,epsf,subfigure,amsbsy,amsmath,amsfonts,float}
\usepackage[latin1]{inputenc}

\begin{document}

\title{EIT and AT: Two Similar but Distinct Phenomena in Two Categories of Three-Level Atomic Systems}

\author{Tony Y. Abi-Salloum}

\affiliation{Physics and Astronomy Department, Widener University,
Chester, Pennsylvania 19013, USA}

\date{\today, {\it Physical Review A}: To be submitted.}

% *****************************************************************************      Abstract
\begin{abstract}
Electromagnetically Induced Transparency (EIT) and Autler-Townes (AT) are two phenomena that could be featured by a variety of three-level atomic systems. The considered phenomena, EIT and AT, are similar ``looking" in the sense that they both are characterized by a reduction in absorption of a weak field in the presence of a stronger field. In this paper, we explicitly set the threshold of separation between EIT and AT in a unified study of four different three-level atomic systems. Two resonances are studied and compared in each case. A comparison of the magnitudes of the resonances reveals two coupling field regimes and two categories of three-level systems.
\end{abstract}

\pacs{PACS numbers: 42.50.Gy}

\maketitle

% ********************************************************************************        Introduction
\section{Introduction}
\label{Introduction}

\begin{figure}[htbp]
\centering
\includegraphics[angle=0,width=14.0cm]{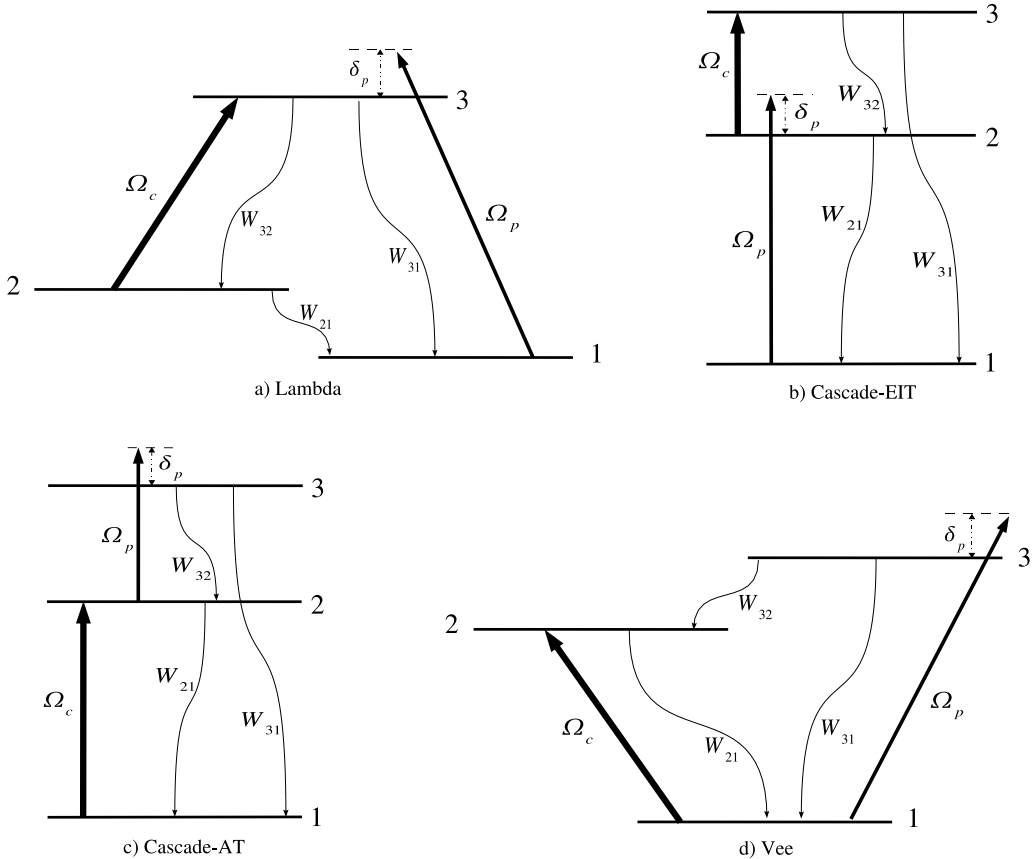}
\caption{$W_{ij}$ is the population decay rate from level i to level j. $\Omega_p$ and $\Omega_c$ denote the Rabi frequencies of the probe and coupling fields respectively. $\delta_p$ is the detuning of the probe field. The coupling field is at resonance}
\label{Fig_1}
\end{figure}

Since the emergence of Electromagnetically Induced Transparency (EIT) in 1990 \cite{Harris:1990} the phenomenon has been associated with destructive interference between two excitation pathways \cite{Harris:1997}. In order to briefly introduce EIT along with its corresponding assumptions and understanding we review first the four three-level atomic systems, Lambda (Fig. \ref{Fig_1}.a) \cite{Li:1995}, Cascade-EIT (Fig. \ref{Fig_1}.b) \cite{Jason:2001}, Cascade-AT (Fig. \ref{Fig_1}.c) \cite{Zhao:1997} and Vee (Fig. \ref{Fig_1}.d) \cite{Vdovic:2007}, which have been always assumed to feature EIT. The Cascade-EIT and Cascade-AT names are adopted in correspondence with our earlier works, ``Interference Between Competing Pathways in the Interaction of Three-Level Ladder Atoms and Radiation" \cite{Tony:2007_Ladder} and ``Phase Dynamics and Interference in EIT" \cite{Tony:2007_Phase-Dynamic}.

%%%%%%%%%%%%%%%%%%%%%%%%%%%%%%%%%
%%%%%%%%%%%%%%%%%%%%%%%%%%%%%%%%%

\begin{figure}[htbp]
\centering
\includegraphics[angle=0,width=10.0cm]{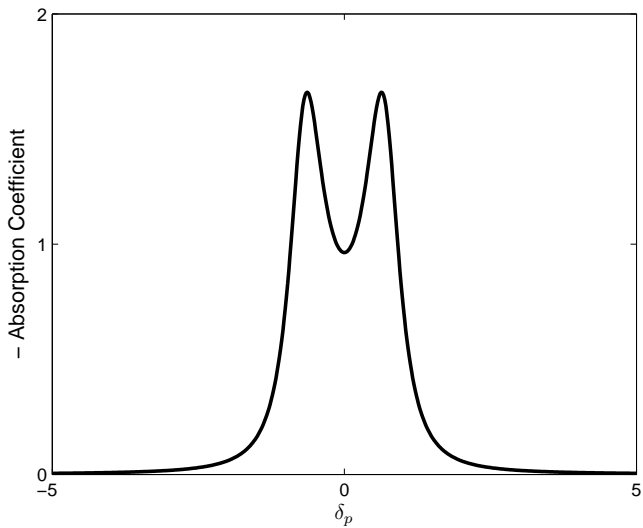}
\caption{The negative absorption coefficient of the probe field function of its detuning. The dip (reduction in absorption) could be a signature of EIT or AT. This is a general plot added for qualitative reasons only.}
\label{Fig_2}
\end{figure}

Each one of the four different three-level systems (Fig. \ref{Fig_1}) is engineered by the action of two fields, one weak field called the probe and one stronger field called the coupling, on two different atomic transitions which share a common level. If we monitor the absorption of the probe field we find (Fig. \ref{Fig_2}) a reduction in absorption (a dip in the absorption line) at resonance ($\delta_p = 0$, the case when the frequency of the probe field matches the atomic transition frequency) where a maximum absorption is expected in the absence of the coupling field. Note that figure \ref{Fig_2} is a general plot added for qualitative understanding only. For the sake of simplicity and clarity of the results we consider the resonance coupling case (the frequency of the coupling field matches the corresponding atomic transition frequency) throughout this work.

%%%%%%%%%%%%%%%%%%%%%%%%%%%%%%%%%
%%%%%%%%%%%%%%%%%%%%%%%%%%%%%%%%%

The surprising reduction in absorption in EIT has always been understood in the literature as a result of destructive interference between two competing excitation pathways \cite{Harris:1997}. These two pathways were explicitly studied and unambiguously presented in the Lambda (Fig. \ref{Fig_1}.a) \cite{Lounis:1992} and Cascade-EIT (Fig. \ref{Fig_1}.b) \cite{Tony:2007_Ladder} cases. We mention in our earlier work \cite{Tony:2007_Ladder} that in the Cascade-AT case, the second resonance (excitation pathway) is negligible compared to the first resonance in the low saturation limit (the coupling Rabi frequency is much weaker than a corresponding atomic polarization decay rate, a point that will be covered in detail later in this paper). Some limitations in the scattering theory \cite{Tony:2007_Ladder} constricts the study of the Cascade-AT model to the low saturation limit, leaving the picture slightly incomplete. In addition to the missing pieces in the Cascade-AT configuration, no work to our knowledge clearly and unambiguously proves the existence of EIT in the Vee system. It is true that dips (reductions) in the absorption lines have been experimentally reported in all four three-level systems but what needs to be clarified is wether the reported results are consequences of EIT or a similar phenomenon known as Autler-Towns splitting (AT) \cite{AT:1955}. The later phenomenon has a signature that looks very similar to that of EIT. Both phenomena display a reduction in absorption where a maximum is expected in the absence of the coupling field (Fig. \ref{Fig_2}). An explicit study of AT was conducted by Cohen-Tannoudji and co-authors in different references \cite{Cohen-Tannoudji:Book, Cohen-Tannoudji:1977-3, Cohen-Tannoudji:Amazing-Light}. The studies where conducted in the secular limit (coupling field much stronger than the polarization decay rates) and showed that the absorption line is made of two Lorentzian-like lines which are located next to each other. We can tell now that the dip in figure \ref{Fig_2} can be interpreted as either a destructive interference between two competing pathways, what is know as EIT, or a gap between two resonances, what is known as AT. It is the goal of this work to study in parallel all four three-level systems (Fig. \ref{Fig_1}) and explicitly clarify when the detected reduction in absorption is a result of EIT and when it is the result of AT.\\

In the next section we use the scattering theory results \cite{Tony:2007_Ladder} to split every one of the four probe absorption spectra (one spectrum for each three-level system) into two terms which we call resonances. The comparison of the two resonances reveals two distinct coupling field regimes and two categories of three-level atomic systems. In section \ref{EIT_AT} one sample of each category is explicitly explored. A study of the two considered configurations leads to conclusions about the existence or absence of EIT and AT in each of the coupling field regimes. Note that a homogenously broadened medium and a resonant coupling field ($\delta_c=0$) cases are considered in this work for the sake of simplicity and clarity of the results. A non-zero coupling detuning and an inhomogeneous doppler broadening may add new physics to the systems but will not change the core of the physical phenomena that are studied in this paper and correspondingly the presented results.

% ********************************************************************************
\section{Two Coupling Field Regimes and Two Categories of Three-Level Systems}
\label{Two Two}

The absorption coefficient of the probe field is derived under the condition that the coupling field is much stronger than the probe field, which is weak enough to get effected by the medium without changing its characteristics. Many techniques can be adopted to derive the explicit form of the absorption coefficient. One of the most favorable techniques that has been used in the literature is the semiclassical technique \cite{Sargent:Book} where the atomic equations of motion are derived then perturbatively solved in steady state for different orders of the probe field. It is not the goal of this paper to cover any of the existing techniques. We consider in this work the spectra that have been derived in different publications. We modify the considered equations of the absorption coefficients, and accordingly the defined variables, to match our three-level systems. The absorption spectra are then analyzed in light of the results of the scattering theory \cite{Tony:2007_Ladder}. We note that the absorption coefficient of the probe field is proportional to the imaginary part of the density matrix element $\rho_{ij}$, where i-j is the atomic transition on which the probe field acts. In this paper we define the needed variables in the following way: Detuning of the probe field $\delta_p = \omega_p - \omega_{ij}$, detuning of the coupling field $\delta_c = \omega_c - \omega_{kl} = 0$ (resonance case), Rabi frequency of the probe field $\Omega_p = 2 \mu_{ij} \mathcal{E}^o_p / \hbar$, Rabi frequency of the coupling field $\Omega_c = 2 \mu_{kl} \mathcal{E}^o_c / \hbar$ and polarization decay rates $\gamma_{mn} = \sum_{t=1}^3\left(W_{mt}+W_{nt}\right)$ where, i-j and k-l are the transitions on which the probe and coupling fields act respectively.\\

In the Lambda case (Fig. \ref{Fig_1}.a) the density matrix element $\rho_{13}$ \cite{Lounis:1992},

\begin{equation}
\rho_{13} \;\propto\;
\dfrac{\delta_p-i\gamma_{12}}{
    |\Omega_c|^2 / 4 - [\delta_p-i\gamma_{13}][\delta_p-i\gamma_{12}]},
\label{rho_12_Lambda}
\end{equation}

is derived as the sum (beside a multiplication factor) of the following two resonances:

\begin{subequations}
\begin{eqnarray}
\rm{1^{st} \; Resonance} &\propto& \dfrac{1}{(Z_{I} - Z_{II})} \times \dfrac{Z_{I}+i\gamma_{12}}{\delta_p -Z_{I}},\label{Resonance_1_Lambda}\\
\rm{2^{nd} \; Resonance} &\propto& - \dfrac{1}{(Z_{I} - Z_{II})} \times \dfrac{Z_{II}+i\gamma_{12}}{\delta_p -Z_{II}},
\label{Resonance_2_Lambda}
\end{eqnarray}
\label{Resonances_Lambda}
\end{subequations}

where,

\begin{subequations}
\begin{eqnarray}
Z_{I} &=& \dfrac{1}{2}\left(-i\gamma_{23} + \sqrt{-(\gamma_{13}-\gamma_{12})^2+|\Omega_c|^2}\right),\\
Z_{II} &=& \dfrac{1}{2}\left(-i\gamma_{23} - \sqrt{-(\gamma_{13}-\gamma_{12})^2+|\Omega_c|^2}\right).
\label{Z_II_Lambda}
\end{eqnarray}
\label{Z_I_Z_II_Lambda}
\end{subequations}

In the Cascade-EIT case (Fig. \ref{Fig_1}.b) the density matrix element of interest is $\rho_{12}$ \cite{Tony:2007_Ladder} given by

\begin{equation}
\rho_{12} \;\propto\;
\dfrac{\delta_p-i\gamma_{13}}{
    |\Omega_c|^2 / 4 - [\delta_p-i\gamma_{12}][\delta_p-i\gamma_{13}]}.
\label{rho_12_Cascade-EIT}
\end{equation}

With the help of the scattering theory \cite{Tony:2007_Ladder} we split the absorption coefficient into the following two resonances:

\begin{subequations}
\begin{eqnarray}
\rm{1^{st} \; Resonance} &\propto& \dfrac{1}{(Z_{I} - Z_{II})} \times \dfrac{Z_{I}+i\gamma_{13}}{\delta_p -Z_{I}},\label{Resonance_1_Cascade-EIT}\\
\rm{2^{nd} \; Resonance} &\propto& - \dfrac{1}{(Z_{I} - Z_{II})} \times \dfrac{Z_{II}+i\gamma_{13}}{\delta_p -Z_{II}},
\label{Resonance_2_Cascade-EIT}
\end{eqnarray}
\label{Resonances_Cascade-EIT}
\end{subequations}

where,

\begin{subequations}
\begin{eqnarray}
Z_{I} &=& \dfrac{1}{2}\left(-i\gamma_{23} + \sqrt{-(\gamma_{12}-\gamma_{13})^2+|\Omega_c|^2}\right),\\
Z_{II} &=& \dfrac{1}{2}\left(-i\gamma_{23} - \sqrt{-(\gamma_{12}-\gamma_{13})^2+|\Omega_c|^2}\right).
\label{Z_II_Cascade-EIT}
\end{eqnarray}
\label{Z_I_Z_II_Cascade_EIT}
\end{subequations}

In light of the work done in the scattering theory and because all four absorption coefficients have similar algebraic forms, the absorption coefficients in the Cascade-AT and Vee cases can as well be written as a sum of two resonances. In the Cascade-AT (Fig. \ref{Fig_1}.c) case the density matrix element $\rho_{23}$ \cite{Tony:2007_Ladder} is given by

\begin{equation}
\rho_{23} \; \propto \; \dfrac{\delta_p -i\gamma_{23}}{
|\Omega_c|^2 / 4-[\delta_p-i\gamma_{23}][\delta_p-i\gamma_{13}]},
\label{rho_23_Cascade-AT}
\end{equation}

$\left( \rm{we \; dropped \; the \; prefactor} \dfrac{|\Omega_c|^2/ 4}{ \gamma_{12}^2 + 2 |\Omega_c|^2 / 4}\right)$ which splits into the two resonances

\begin{subequations}
\begin{eqnarray}
\rm{1^{st} \; Resonance} &\propto& \dfrac{1}{(Z_{I} - Z_{II})} \times \dfrac{Z_{I}+i\gamma_{23}}{\delta_p -Z_{I}}\label{Resonance_1_Cascade-AT},\\
\rm{2^{nd} \; Resonance} &\propto& - \dfrac{1}{(Z_{I} - Z_{II})} \times \dfrac{Z_{II}+i\gamma_{23}}{\delta_p -Z_{II}},
\label{Resonance_2_Cascade-AT}
\end{eqnarray}
\label{Resonances_Cascade-AT}
\end{subequations}

where,

\begin{subequations}
\begin{eqnarray}
Z_{I} &=& \dfrac{1}{2}\left(-i(\gamma_{23}+\gamma_{13}) + \sqrt{-\gamma_{12}^2+|\Omega_c|^2}\right),\\
Z_{II} &=& \dfrac{1}{2}\left(-i(\gamma_{23}+\gamma_{13}) - \sqrt{-\gamma_{12}^2+|\Omega_c|^2}\right).
\label{Z_II_Cascade-AT}
\end{eqnarray}
\label{Z_I_Z_II_Cascade_AT}
\end{subequations}

In a similar manner, the density matrix element $\rho_{13}$ \cite{Lazoudis:2005} is given in the Vee case (Fig. \ref{Fig_1}.d) by

\begin{equation}
\rho_{13} \;\propto\;
\dfrac{(\delta_p-i\gamma_{13})
     \; \dfrac{|\Omega_c|^2}{ 4\gamma_{12}^2} \;+ (\delta_p-i\gamma_{23})} {|\Omega_c|^2 /4 - [\delta_p-i\gamma_{23}][\delta_p-i\gamma_{13}]},
\label{rho_13_Vee}
\end{equation}

$\left( \rm{we \; dropped \; the \; prefactor} \dfrac{1}{1+\dfrac{|\Omega_c|^2}{ 2\gamma_{12}^2}}\right)$ which can be re-written as the sum of the two terms

\begin{subequations}
\begin{eqnarray}
\rm{1^{st} \; Resonance} &\propto& \dfrac{1}{(Z_{I} - Z_{II})} \times \dfrac{Z_{I}\left(1+\dfrac{|\Omega_c|^2}{ 4\gamma_{12}^2}\right)+i\left(\gamma_{23}+\gamma_{13}\dfrac{|\Omega_c|^2}{ 4\gamma_{12}^2}\right)}{\delta_p -Z_{I}}\label{Resonance_1_Vee},\\
\rm{2^{nd} \; Resonance} &\propto& - \dfrac{1}{(Z_{I} - Z_{II})} \times \dfrac{Z_{II}\left(1+\dfrac{|\Omega_c|^2}{ 4\gamma_{12}^2}\right)+i\left(\gamma_{23}+\gamma_{13}\dfrac{|\Omega_c|^2}{ 4\gamma_{12}^2}\right)}{\delta_p -Z_{II}},
\label{Resonance_2_Vee}
\end{eqnarray}
\label{Resonances_Vee}
\end{subequations}

where,

\begin{subequations}
\begin{eqnarray}
Z_{I} &=& \dfrac{1}{2}\left(-i(\gamma_{23}+\gamma_{13}) + \sqrt{-\gamma_{12}^2+|\Omega_c|^2}\right),\\
Z_{II} &=& \dfrac{1}{2}\left(-i(\gamma_{23}+\gamma_{13}) - \sqrt{-\gamma_{12}^2+|\Omega_c|^2}\right).
\label{Z_II_Vee}
\end{eqnarray}
\label{Z_I_Z_II_Vee}
\end{subequations}

\begin{figure}[htbp]
\centering
\includegraphics[angle=0,width=12.0cm]{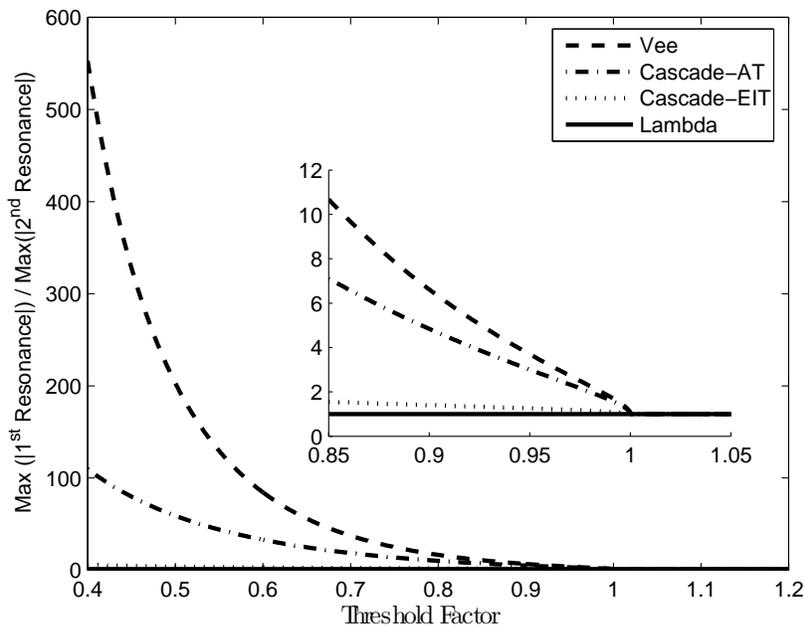}
\caption{Relative magnitudes of the two resonances (Eqs. \ref{Resonances_Lambda} (Lambda), \ref{Resonances_Cascade-EIT} (Cascade-EIT), \ref{Resonances_Cascade-AT} (Cascade-AT), \ref{Resonances_Vee} (Vee)) function of the threshold factor, $\Omega_c$/threshold (Eqs. \ref{Thresholds} a-d)}
\label{Fig_3}
\end{figure}

We now know that all four absorption coefficients split into two resonances and that in the two Cascade-EIT \cite{Tony:2007_Ladder} and Lambda \cite{Lounis:1992} cases the two resonances interfere in the weak coupling field regime (coupling Rabi frequency less than a corresponding polarization decay rate). Based on the previously stated fact we may assume that the same type of interference which exists in the Cascade-EIT and Lambda configurations must exist in the Cascade-AT and Vee cases. We showed in our previous work \cite{Tony:2007_Ladder} that in the low saturation limit (very weak coupling field regime) no interference can exist in the Cascade-AT case. We plot in figure \ref{Fig_3} the ratio of the maxima of the absolute values of the resonances, which reflects the relative strengths between the two resonances, versus what we call the ``threshold factor" which we define as $\Omega_c$/threshold. The threshold is the value of $\Omega_c$ which separates between the weak and strong coupling field regimes. The threshold varies from one system to another. A theocratical study (not included in this work) of the resonances shows that the threshold is equal to the polarization decay rates which are under the square roots in the $Z_I$ and $Z_{II}$ complex numbers. The study of these complex numbers leads to the following thresholds in the four different systems:

\begin{subequations}
\begin{eqnarray}
\rm{Lambda} &:& \gamma_{13}-\gamma_{12},\\
\rm{Cascade-EIT} &:& \gamma_{12}-\gamma_{13},\\
\rm{Cascade-AT} &:& \gamma_{12},\\
\rm{Vee} &:& \gamma_{12}.
\end{eqnarray}
\label{Thresholds}
\end{subequations}

Note that the polarization decay rates have different values in the different three-level systems. We consistently use for the theocratical modeling in this paper the following scaled values of the population decay rates $W_{ij}$ (spontaneous decay from level i to level j): Cascade-EIT and Cascade-AT ($W_{21}=1$, $W_{32}=0.2$, $W_{31}=0.001$); Lambda ($W_{31}=1$, $W_{32}=0.9$, $W_{21}=0.001$); Vee ($W_{31}=1$, $W_{21}=0.9$, $W_{32}=0.001$). It is true that the study of the different three-level configurations function of a common coupling field strength is not fair because of the different decay rates that the fields are affected by. For this reason we unify the following study across the different configurations by plotting the ratio of the magnitudes of the resonances versus the threshold factor ($\Omega_c$/threshold).

%%%%%%%%%%%%%%%%%%%%%%%%%%%%%%%%%
%%%%%%%%%%%%%%%%%%%%%%%%%%%%%%%%%

Figure \ref{Fig_3} shows clearly the existence of two regimes and two categories of three-level systems. In the strong coupling field regime, threshold factor $>$ 1 ($\Omega_c >$ threshold), the two resonances have exactly equal magnitudes. The magnitudes of the resonances vary as the strengths of the coupling field varies in the strong field regime but the ratio of the magnitudes remains equal to 1. Once the coupling field Rabi frequency becomes less than the threshold value (threshold factor $<$ 1) the four three-level systems show two completely different behaviors. In the Lambda and Cascade-EIT cases the two resonances remain competitive keeping a ratio that is very close to 1 (this characteristic is easily seen in the insert of figure \ref{Fig_3}). The surprising results are in the Cascade-AT and Vee configurations where one resonance dominates the other as the coupling field becomes weaker. Figure \ref{Fig_3} has the clear evidence that two different coupling field regimes must be distinguished and that two categories of three-level systems must be considered. We know so far from this work that any possible interference effects that exist between the two resonances in the Cascade-AT and Vee cases is going to be negligible in the low saturation limit (very weak coupling field regime). The later conclusion is inline with our study \cite{Tony:2007_Ladder} of the Cascade-AT system where we state that only one resonance has to be retained in the low saturation limit.

% ********************************************************************************
\section{Distinction Between EIT and AT}
\label{EIT_AT}

The study of the relative magnitudes of the resonances (Sec. \ref{Two Two}) revealed two categories of three-level systems. We study in this section one system from each category for further understanding of the differences between these systems and distinction between the EIT and AT phenomena. The two configurations which we are going to study in depth are the Lambda (Fig. \ref{Fig_1}.a) and Vee (Fig. \ref{Fig_1}.d) which happen to be the most significant cases in their categories.\\

%%%%%%%%%%%%%%%%%%%%%%%%%%%%%%%%%
\begin{figure}[htbp]
\centering
\includegraphics[angle=0,width=9cm]{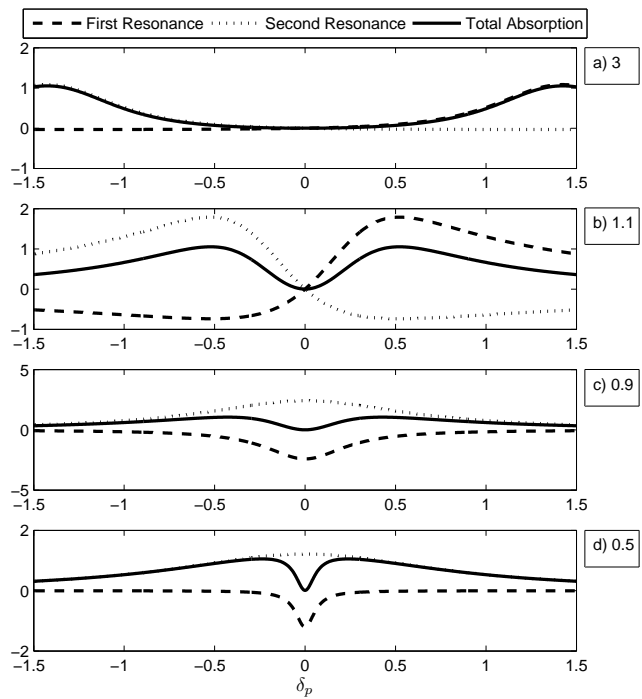}
\caption{``Evolution" of the resonances function of the decreasing (subplots a-d) threshold factor in the Lambda case (Fig. \ref{Fig_1}.a). Each subplot includes the first resonance, second resonance, and total absorption of the probe field function of its detuning.}
\label{Fig_4}
\end{figure}
%%%%%%%%%%%%%%%%%%%%%%%%%%%%%%%%%

Figure \ref{Fig_4} shows the ``evolution" of the two resonances function of the threshold factor in the Lambda case. Each subplot displays the two resonances separately and combined as the total absorption coefficient function of the detuning of the probe (a plot which is similar to figure \ref{Fig_2}). Subplots a, b, c and d show the evolution of the resonances and total absorption as we decrease the threshold factor. Subplot \ref{Fig_4}.a is a strong coupling field regime case. We can see that the total absorption is made of two peaks, each of which is a resonance. The observed dip can be interpreted as a gap between the two resonances. This reduction in absorption is a characteristic of the AT effect as explained by Cohen-Tannoudji and co-authors \cite{Cohen-Tannoudji:Book}. Subplot \ref{Fig_4}.b is a strong coupling field regime case as well but it is right above the threshold. We can see in this subplot that the dip is still a consequence of a gap between the two resonances even though the two resonances overlap. Once the threshold factor is less than 1 ($\Omega_c <$ threshold) we realize two crucial changes (subplot \ref{Fig_4}.c). The two resonances totally overlap (peak on peak) and one resonance becomes totally negative. In this case, weak coupling field regime, the dip is a result of a destructive interference between the two resonances. The detected reduction in absorption is an ``imprint" of one resonance into the other. This destructive interference, which is a signature of EIT, becomes more apparent in subplot \ref{Fig_4}.d where the coupling field is weaker than what it was in the previous case (subplot \ref{Fig_4}.c). This set of plots (Fig. \ref{Fig_4}) unambiguously shows that EIT and AT are two similar looking but distinct phenomena. Even though the two phenomena are consequences of two resonances, the detected reduction in absorption is a result of two different pieces of physics. This study of the Lambda system clarifies the existence of AT and EIT in the strong and weak coupling field regimes respectively in the Lambda and Cascade-EIT configurations.

%%%%%%%%%%%%%%%%%%%%%%%%%%%%%%%%%
\begin{figure}[htbp]
\centering
\includegraphics[angle=0,width=9cm]{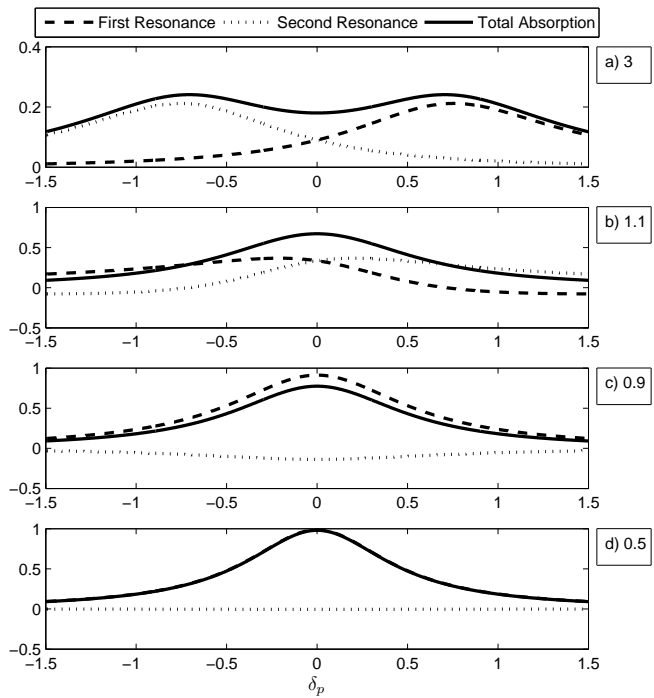}
\caption{``Evolution" of the resonances function of the decreasing (subplots a-d) threshold factor in the Vee case (Fig. \ref{Fig_1}.d). Each subplot includes the first resonance, second resonance, and total absorption of the probe field function of its detuning.}
\label{Fig_5}
\end{figure}
%%%%%%%%%%%%%%%%%%%%%%%%%%%%%%%%%

The same way we previously studied the Lambda system we now study the Vee system. Figure \ref{Fig_5} is a set of subplots similar (same threshold factors) to the ones presented in figure \ref{Fig_4}. The AT feature, a gap between two neighboring resonances, is the same in the Vee case (subplot \ref{Fig_5}.a) as it is in the Lambda in the strong coupling field regime. Even though the two resonances overlap with the peaks one next to the other, the dip disappears in subplot \ref{Fig_5}.b where the coupling Rabi frequency is right above the threshold value. This disappearance of the dip has to do with the fact that the resonances become shallower as the coupling Rabi frequency decreases. In order to study the EIT phenomenon in the Vee system we consider the weak coupling field regime (subplots \ref{Fig_5}.c-d). In consistence with the Lambda case the two resonances overlap peak on peak. The difference here is that one of the resonances becomes very flat and starts vanishing as the coupling field becomes weaker (expected from the study of figure \ref{Fig_3}). These dramatic changes in one of the resonances cut off any possibility of seeing a dip in the absorption line. We can easily notice in subplot \ref{Fig_5}.d that one resonance is completely negligible while the other perfectly overlaps with the total absorption line. In other words, the absorption spectrum in the low saturation limit (very weak coupling field) is a result of only one resonance. This conclusion is inline with our previous studies \cite{Tony:2007_Ladder} presented in the Cascade-AT case. The evolution of the resonances in the Vee case verifies the absence of EIT in the Vee (Fig. \ref{Fig_1}.d) and Cascade-AT (Fig. \ref{Fig_1}.c) configurations.

% ********************************************************************************
\section{Conclusion}
\label{Conclusion}

We explicitly studied in this paper the four three-level atomic systems (Fig. \ref{Fig_1}) that feature EIT and/or AT. The different studies were conducted under the only limit of coupling resonance ($\delta_c=0$). We considered a homogenously broadened medium and a resonant coupling field ($\delta_c=0$) for the sake of simplicity and clarity of the results. The drawn conclusions are based on the study of two resonances for each system that add up to the exact total absorption coefficient. The four schemes were compared in light of the variation of the threshold factor ($\Omega_c$/threshold) which unifies the studies across the different systems. The value 1 ($\Omega_c$ = threshold) of the threshold factor is clearly a point of separation between the strong ($\Omega_c  >$ threshold) and weak ($\Omega_c  <$ threshold) coupling field regimes as shown in figure \ref{Fig_3}. This fact verifies the values of the thresholds (Eqs. \ref{Thresholds}) and the unification of the conducted studies.\\

In the strong coupling field regime (threshold factor $>$ 1) an exact equality between the absolute magnitudes of the resonances is common across the four systems (Fig. \ref{Fig_3}). The peaks of the resonances are located next to each other with overlapping wings (subplots \ref{Fig_4}.a and \ref{Fig_5}.a). The observed reduction in absorption is a consequence of a gap between the resonances, which is a characteristic of AT splitting. In summary, the AT splitting (two resonances with a gap in between) is commonly observed in the strong coupling field regime ($\Omega_c \; >$ threshold) in the four different three-level systems (Fig. \ref{Fig_1}).\\

In the weak coupling field regime (threshold factor $<$ 1) the four three level systems split into two categories (Fig. \ref{Fig_3}). The Lambda and Cascade-EIT configurations continue to feature two highly competitive resonances unlike the case in the Casade-AT and Vee systems where one resonance dominates the other as the threshold factor decreases. This characteristic which sets the two categories apart is most apparent in subplots \ref{Fig_4}.d and \ref{Fig_5}.d. A dip in the absorption line can be seen (Fig. \ref{Fig_4}.c-d) in the Lambda and Cascade-EIT systems which belong to the same category of three-level systems. This dip is a result of destructive interference between the two competing pathways, which is a characteristic of EIT. This later dip is absent in the second category of three-level systems which includes the Cascade-AT and Vee configurations (Fig. \ref{Fig_4}.c-d). As the threshold factor decreases one resonance dominates the other (Fig. \ref{Fig_5}.c-d) in the the Cascade-AT and Vee systems. In these later cases, Cascade-AT and Vee, and in the low saturation limit (very weak coupling field such as the case in subplot \ref{Fig_5}.d) the absorption spectrum becomes equal to the contribution of only one resonance in the presence of the other negligible resonance. In summary, an EIT type of a dip which is a result of destructive interference can only be observed in the Lambda (Fig. \ref{Fig_1}.b) and Cascade-EIT (Fig. \ref{Fig_1}.a) systems but never in the Cascade-AT (Fig. \ref{Fig_1}.c) and Vee (Fig. \ref{Fig_1}.d) systems.\\

% *********************************************************************************************************

\section{Acknowledgments}

We thank Dr. Frank Narducci for the inspiring discussions. We also sincerely thank Professor Mark Havey for his careful reading of the manuscript and for his constructive feedback.

% *************************************************************************************      Bibliography

\end{document}